\documentclass[a4paper]{spie}  %>>> use this instead for A4 paper
\usepackage[]{graphicx}
\usepackage{amssymb,amsmath,psfrag,url}

\newcommand{\MyField}[1]{{\bf{#1}}}

\newcommand{\Bolivarallee}{Boliva\hspace{-0.1mm}r\hspace{0.15mm}a\hspace{-0.1mm}llee}

\title{Finite element method for accurate 3D simulation of plasmonic waveguides}

\author{
Sven Burger,\supit{\,ab}
Lin Zschiedrich,\supit{\,ab}
Jan Pomplun,\supit{\,ab}
Frank Schmidt\supit{\,ab}
\skiplinehalf
\supit{a}
Zuse Institute Berlin (ZIB),
Takustra{\ss}e 7,
D\,--\,14\,195 Berlin,
Germany
\smallskip\\
\supit{b}
JCMwave GmbH,
\Bolivarallee\ 22, 
D\,--\,14\,050 Berlin,
Germany
}
\authorinfo{
Corresponding author: S. Burger\\
URL: http://www.zib.de/Numerik/NanoOptics/\\
URL: http://www.jcmwave.com\\
Email: burger@zib.de
}

\begin{document}
\maketitle
%\today
%%%%%%%%%%%%%%%%%%%%%%%%%%%%%%%%%%%%%%%%%%%%%%%%%%%%%%%%%%%%% 
%% SPIE Copyright form 
\noindent
This paper will be published in Proc.~SPIE Vol. {\bf 7604}
(2010) 76040F,  
({\it Integrated Optics: Devices, Materials, and Technologies XIV, Jean-Emmanuel Broquin; Christoph M. Greiner, Editors})
and is made available 
as an electronic preprint with permission of SPIE. 
One print or electronic copy may be made for personal use only. 
Systematic or multiple reproduction, distribution to multiple 
locations via electronic or other means, duplication of any 
material in this paper for a fee or for commercial purposes, 
or modification of the content of the paper are prohibited.
%%%%%%%%%%%%%%%%%%%%%%%%%%%%%%%%%%%%%%%%%%%%%%%%%%%%%%%%%%%%% 

\begin{abstract}
Optical properties of hybrid plasmonic waveguides and of low-Q cavities,  
formed by waveguides of finite length are investigated numerically. 
These structures are of interest as building-blocks of plasmon lasers. 
We use a time-harmonic finite-element package including a propagation-mode solver, 
a resonance-mode solver and a scattering solver for studying various properties 
of the system. Numerical convergence of all used methods is demonstrated. 
\end{abstract}


\begin{figure}[b]
\begin{center}
  \includegraphics[width=.32\textwidth]{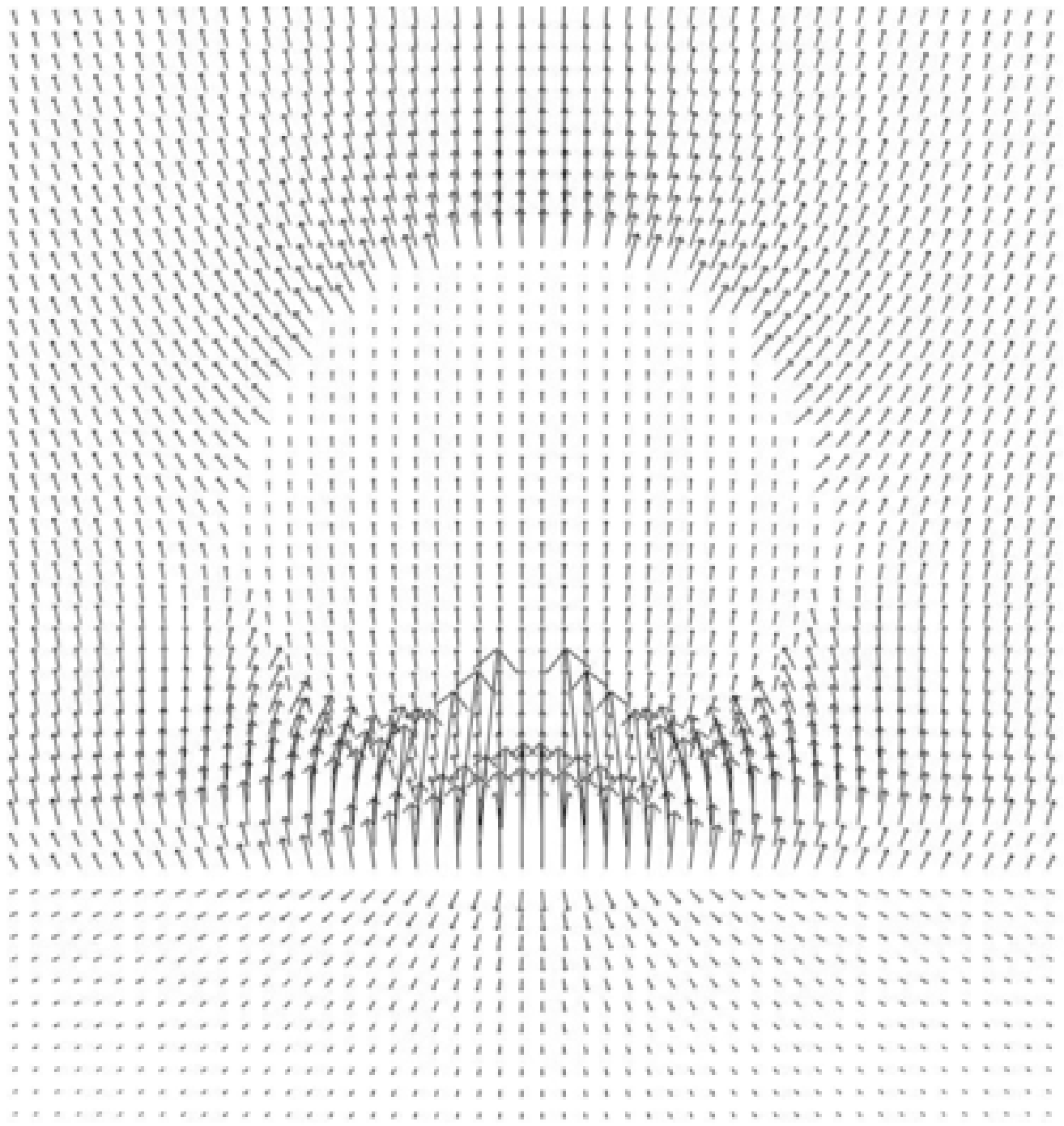}
  \includegraphics[width=.32\textwidth]{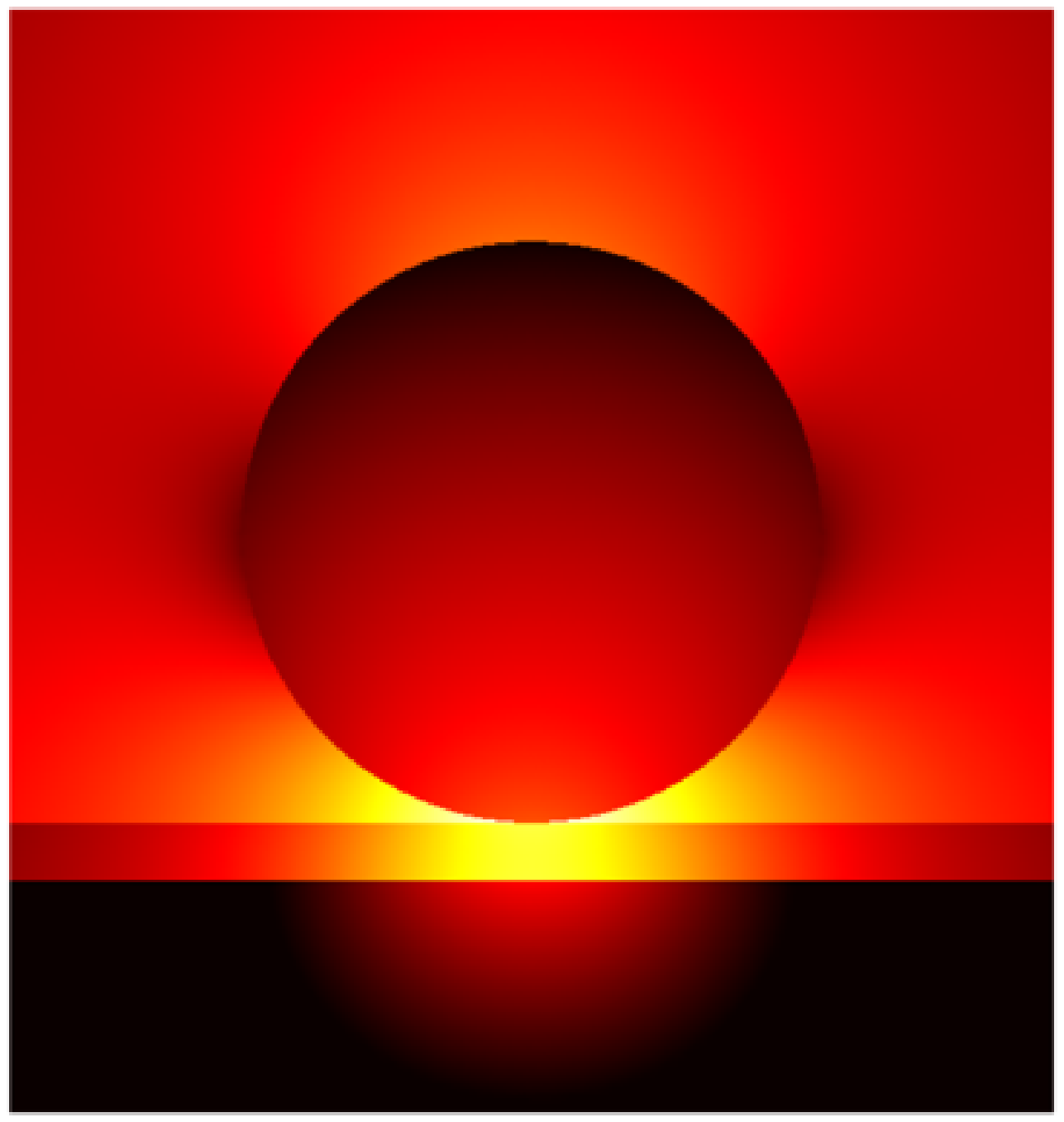}
  \caption{
Visualization of a plasmonic waveguide mode at wavelength $\lambda_0=489\,$nm. 
Left: Projection of the vectorial electric field onto the $xy$-plane. 
Right: Logarithm of the magnitude of the $y$ component of the electric field
(black/dark: low intensity, yellow/bright: high intensity).
The total field of view has a size of 90\,nm$\times$95\,nm.
}
\label{fig_wg_mode}
\end{center}
\end{figure}
\section{Introduction}
Recently, {\it plasmon lasers}, sometimes also called {\it nanolasers} or {\it spasers} 
(surface plasmon lasers)~\cite{Bergman2003prl}, 
have been realized in different systems. 
Essentially, in these devices light can be confined at deep subwavelength scale. 
This opens new prospects for photonic integration and applications like biosensing on a nanoscale~\cite{Anker2008nm}.
Approaches for light confinement include 
nanosphere arrangements~\cite{Noginov2009nature},
rectangular semiconductor nanopillars encapsulated in silver~\cite{Hill2009oe}, 
semiconductor nanowires on top of a silver substrate~\cite{Oulton2009n}, 
bowtie-shaped silver nanoparticle cavities~\cite{Chang2008oe}, and others.
In some of the devices, plasmon-based waveguiding structures play a crucial role. 
Plasmon waveguides incorporating metal and several different dielectric 
materials are able to confine the light field in two dimensions to a very small spot size 
(see typical mode field distribution in Figure~\ref{fig_wg_mode})
at relatively low propagation loss~\cite{Oulton2008njp,Oulton2008np}. 
To distinguish these waveguides 
from simpler approaches consisting of only 
metal and a single dielectric material they are also called {\it hybrid} plasmonic 
waveguides~\cite{Oulton2008np}.

Fast and accurate 3D Maxwell solvers are needed 
for designing structural parameters of such plasmonic waveguides. 
Due to the multi-scale nature of the corresponding field distributions, 
accurate computation of the properties of such devices 
can be numerically challenging. 
We have developed finite-element method (FEM) based solvers for the 
Maxwell eigenvalue and for the Maxwell scattering problems. 
The method is based on higher order vectorial elements, 
adaptive unstructured grids, and on a rigorous treatment of 
transparent boundaries. The method has been applied to  
plasmonic devices like plasmonic antennas, gratings and waveguides~\cite{Hoffmann2009a,Lockau2009a,Tyagi2008oe}.
Here we investigate plasmonic waveguides and cavities in layouts similar to the devices used in 
experiments by the group of X.~Zhang~\cite{Oulton2009n}. 

\begin{figure}[h]
\centering
\psfrag{l1}{\sffamily $L$}
\psfrag{d1}{\sffamily $h_1$}
\psfrag{d2}{\sffamily $D$}
\psfrag{x}{\sffamily $x$}
\psfrag{y}{\sffamily $y$}
\psfrag{z}{\sffamily $z$}
\fbox{\hspace{0.5cm} \includegraphics[width=0.45\textwidth]{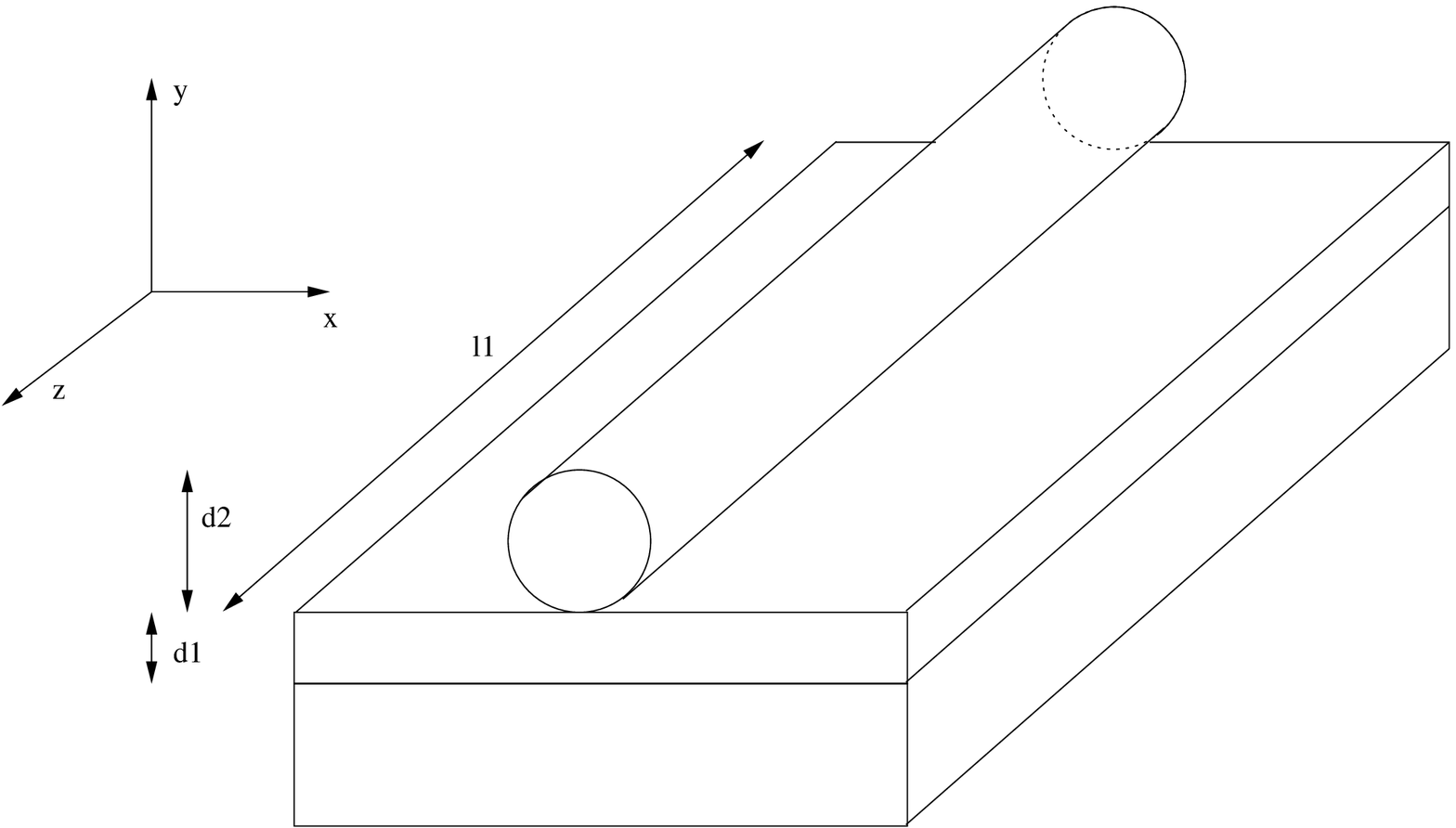}\hspace{0.5cm}}
      \includegraphics[width=.45\textwidth]{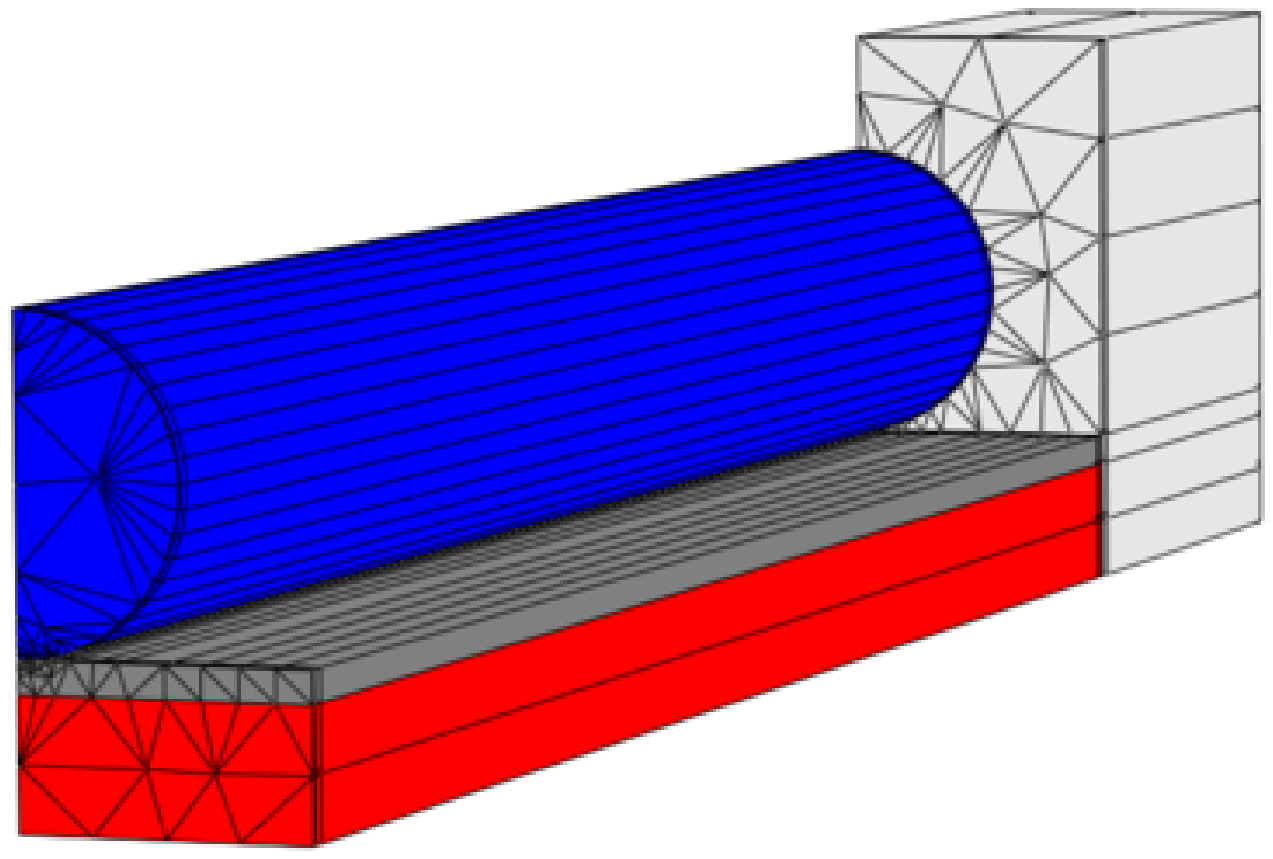}
\caption{Left: Schematic of the setup: CdS nanowire of diameter $D$ and length $L$ placed on a thin layer of 
MgF$_2$ ($h_1=5\,$nm) on a silver layer (subspace). 
Right:
Mesh of the geometry of a plasmonic waveguide 
with finite length $L=374\,$nm in a 3D setup.
}
\label{fig_schema_pl}
\end{figure}

\section{Investigated Setup}
\label{section_setup}

The investigated setup consists of a cadmium sulphide (CdS) semiconductor nanowire 
placed on a thin layer of (insulating) magnesium fluoride (MgF$_2$) on a silver (Ag) 
surface. 
Figure~\ref{fig_schema_pl} (left) shows the schematic of such a device. 
The cadmium sulphide nanowire with diameter $D$ extends in $z$-direction 
over a length $L$. 
The parameters chosen in this numerical study are $D=50$\,nm, $h_1=5\,$nm, 
$L$ of up to few $\mu$m.
The CdS material is pumped optically and lases at a wavelength of $\lambda=489\,$nm. 
However, due to the small size of the structure the supported propagating modes are concentrated 
mainly in the MgF$_2$ region below the nanowire, with some overlap to the amplifying CdS material. 
In the $xy$-cross-section these modes are concentrated at deep subwavelength scale due to the 
plasmonic nature of the excitation at the Ag -- MgF$_2$ interface. 
The chosen geometrical parameters correspond to experimentally realized designs~\cite{Oulton2009n}. 
In our numerical study of optical properties of the plasmonic waveguide design we assume the 
following relative permittivities: 
metal layer: $\varepsilon_r = -9.008+0.3i$,
insulating layer: $\varepsilon_r = 1.9047$,
nanowire: $\varepsilon_r = 6.3$, 
air: $\varepsilon_r = 1$.
Please note that here we study only the passive optical properties of  
waveguide and cavity (assumption of a real permittivity of the nanowire).

\begin{figure}[h]
\begin{center}
  \includegraphics[width=.32\textwidth]{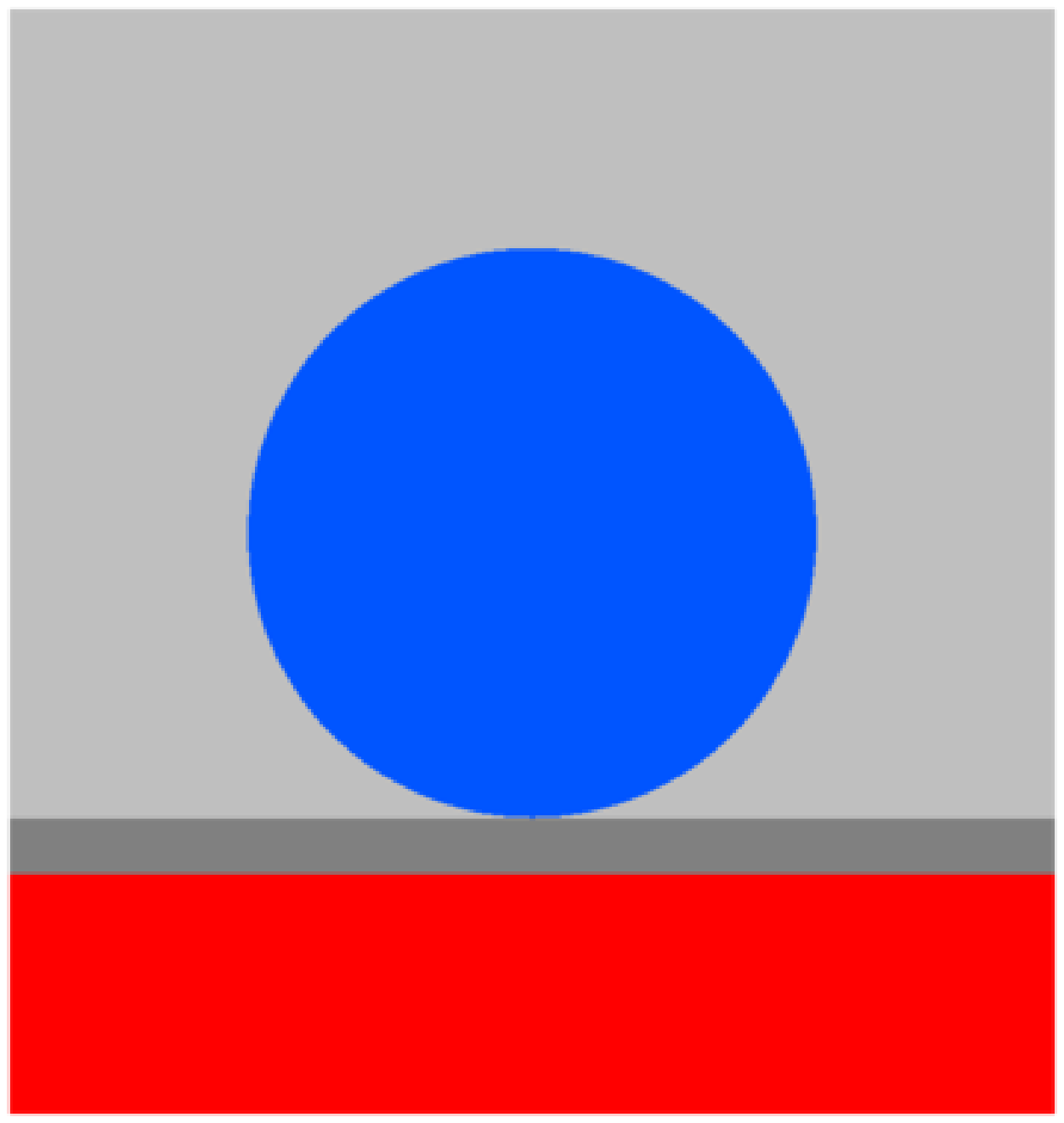}
  \includegraphics[width=.32\textwidth]{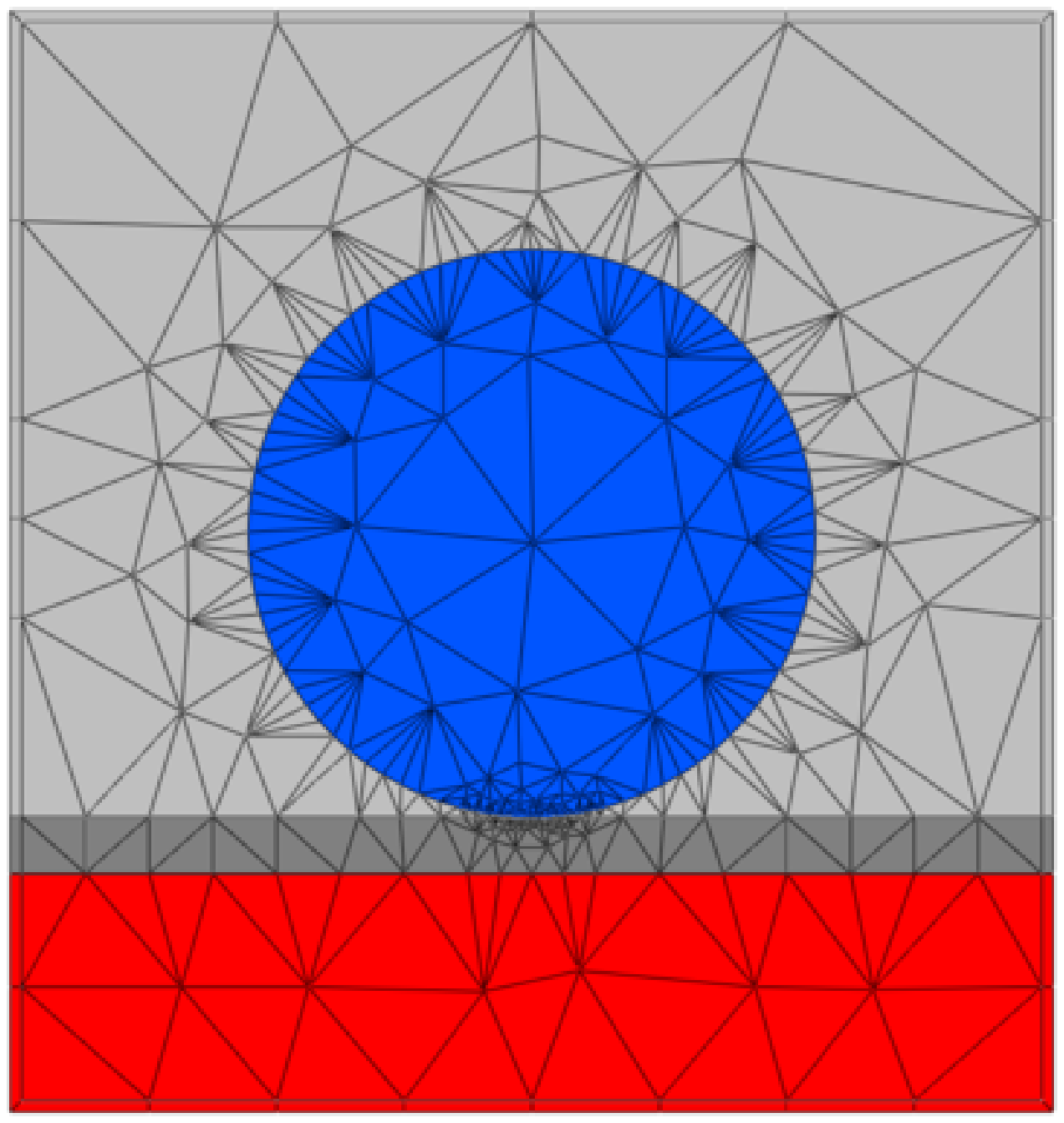}
  \includegraphics[width=.32\textwidth]{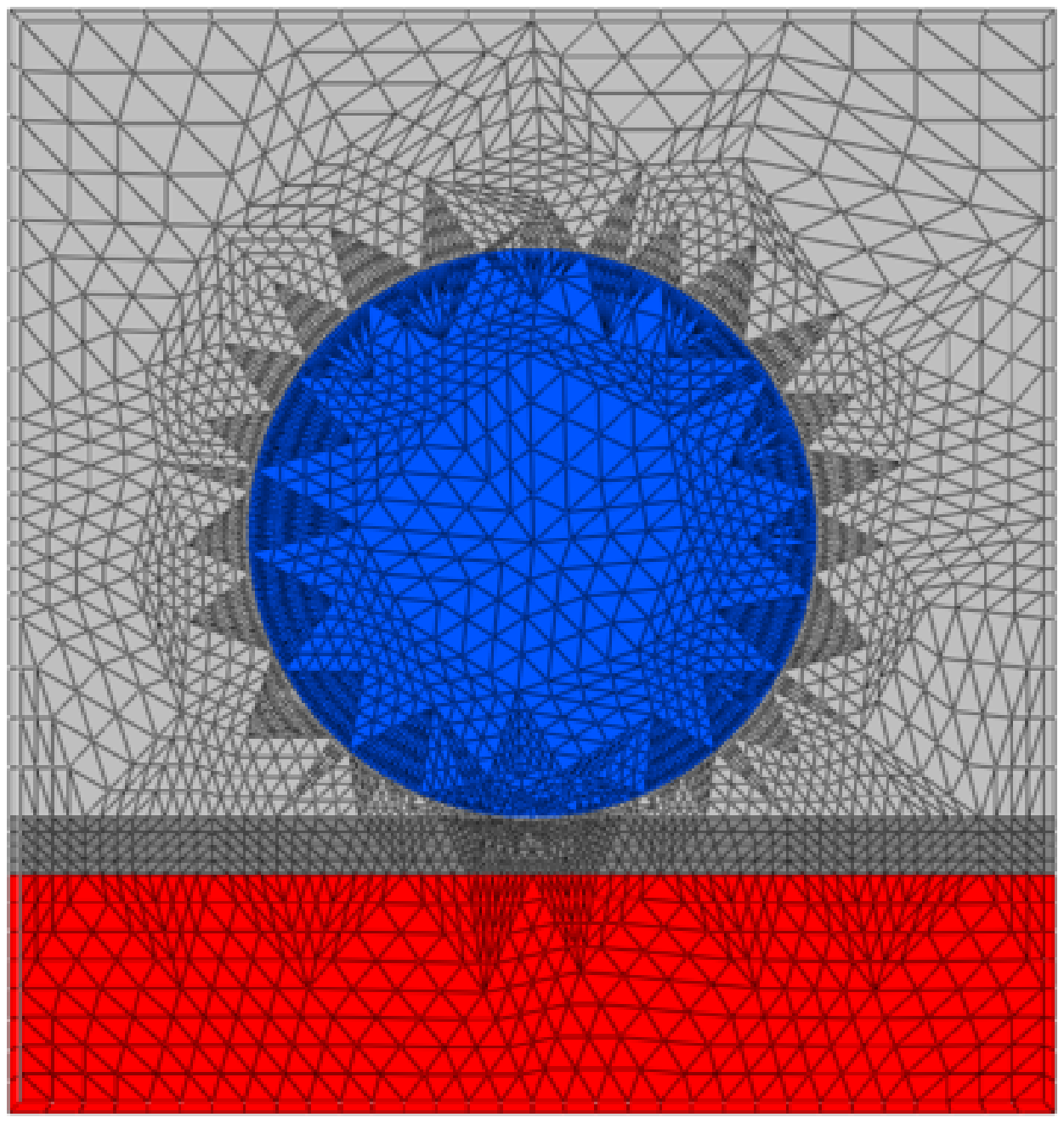}
  \caption{
Left: Plasmonic waveguide geometry ($xy$-cross-section through 
the geometry depicted in Fig.~\ref{fig_schema_pl}): 
CdS nanowire (blue), MgF$_2$ layer (dark grey), Ag (red), surrounded by 
air (light grey). 
Center: Mesh of the geometry for the FEM discretization. 
Right: Adaptively refined mesh (two refinement steps).  
}
\label{fig_geo_grid}
\end{center}
\end{figure}

The 3D geometry and corresponding mesh 
of a plasmonic waveguide 
with finite length $L$ are shown in Figure~\ref{fig_schema_pl}.
The geometry is discretized using prismatoidal elements. 
The subgridding of the 3D mesh in $z$-direction is not shown here.  
For light scattering simulations, a source field which typically is a 2D 
waveguide mode can be coupled into the computational domain at the domain boundaries. 
For eigenvalue computations modelling the resonance of the cavity, no source 
fields are applied.  
Please note that due to a mirror symmetry of the geometry, 
the computational domain can be reduced to the half space $x\ge 0$.
Figure~\ref{fig_geo_grid} shows the geometry of the plasmonic waveguide 
$xy$-cross-section
and the finite-element mesh discretizing 
the geometry. 
Here, the geometry is assumed to be invariant in $z$-direction; MgF$_2$,  Ag
and air regions are modeled as extending to infinity in the $xy$-plane. 
The computational domain on which the mode field is 
computed is discretized with triangular elements while the surrounding 
exterior domain on which purely outgoing fields are assumed is discretized 
with quadrilateral elements. 
The more complex subgridding of the exterior domain for the adaptive 
realization of PMLs is not shown here~\cite{Zschiedrich2006pml}. 

\section{Numerical Results}

\subsection{Guided modes in a plasmonic waveguide}
\label{section_2d}

\begin{figure}[t]
\psfrag{N}{\sffamily $N$}
\psfrag{CPU time [min]}{\sffamily CPU time [min]}
\psfrag{Real(beta)}{\sffamily $\Re (n_\mathrm{eff})$}
\psfrag{Imag(beta)}{\sffamily $\Im (n_\mathrm{eff})$}
\psfrag{Relative error}{\sffamily $\Delta\Re(n_\mathrm{eff}) / \Delta\Im(n_\mathrm{eff})$}
\begin{center}
  \includegraphics[width=.48\textwidth]{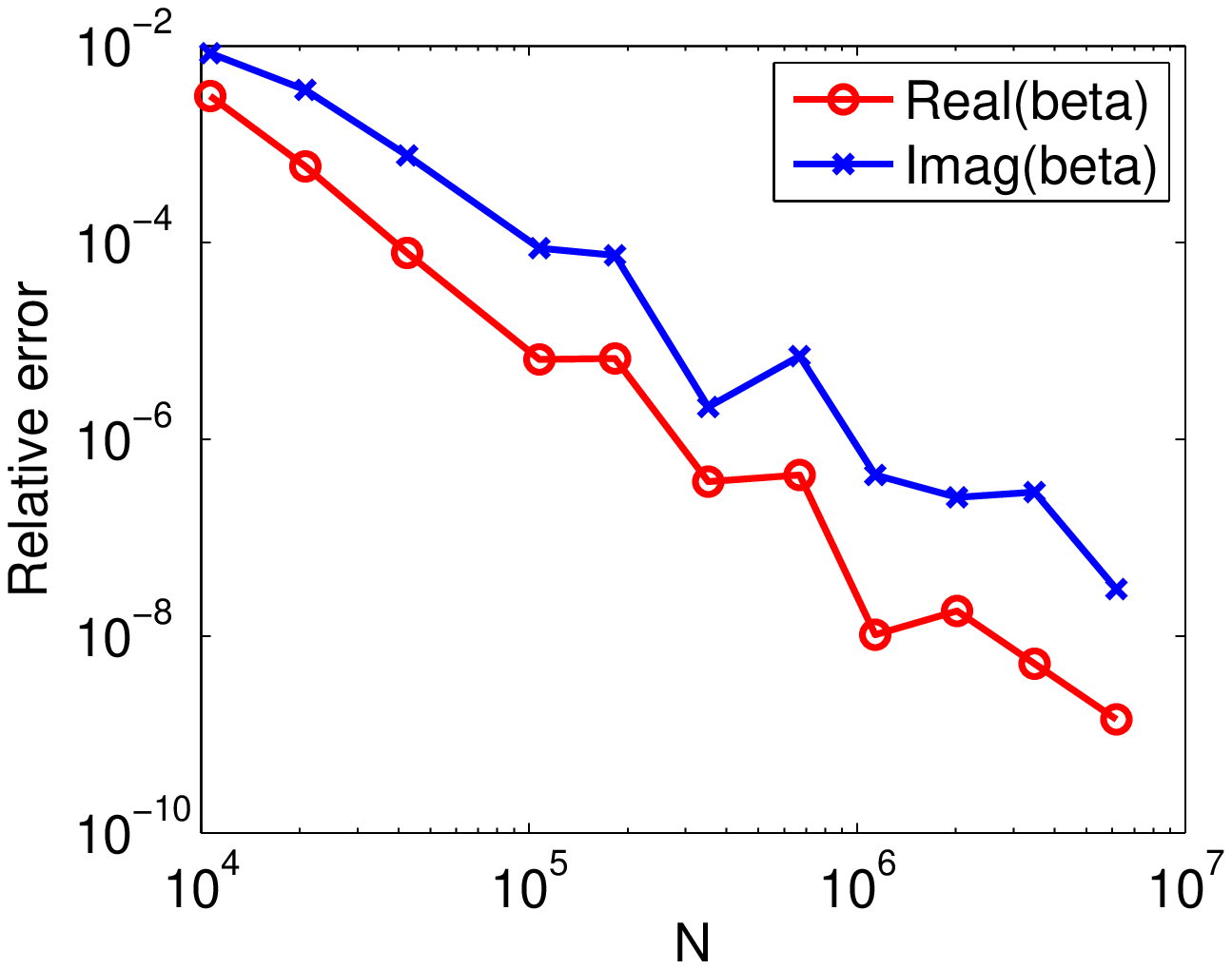}
  \includegraphics[width=.48\textwidth]{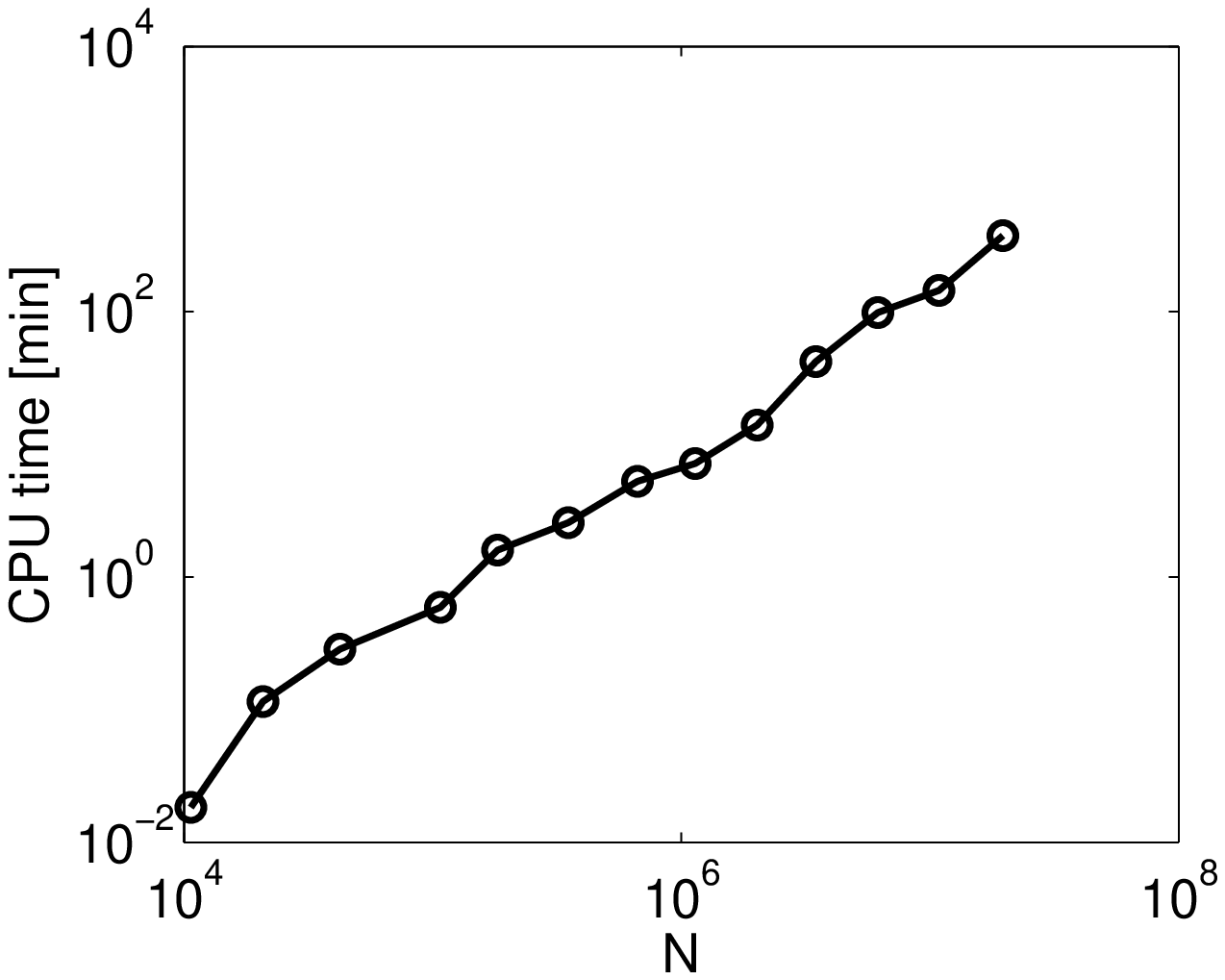}
  \caption{
Computation of propagating modes in a plasmonic waveguide. 
Left: Convergence of the relative error of the real and imaginary parts 
of the eigenvalue (propagation constant) with number of unknowns of 
the FEM problem, $N$. 
Right: Computational effort in terms of cpu-time with number of unknowns 
for the same data set. 
}
\label{fig_wg_mode_conv}
\end{center}
\end{figure}

The geometry of the plasmonic waveguide is invariant $z$-direction, {\it cf.} Figure~\ref{fig_schema_pl}. 
A propagating mode is a solution to the time harmonic Maxwell's equations with frequency $\omega$, 
which exhibits a harmonic dependency in $z$-direction:
$$
\MyField{E}  =  \MyField{E}_{\mathrm{pm}}(x, y)\exp \left(ik_{z}z\right).
$$
$\MyField{E}_{\mathrm{pm}}(x, y)$ is the electric propagation mode on the 2D cross section
and the parameter $k_{z}$ is called propagation constant. 
The effective refractive index $n_{\mathrm{eff}}$ is defined as 
$n_{\mathrm{eff}} = k_z/k_0 \quad \mbox{with} \quad k_0 = 2\pi/\lambda_0$,
where $\lambda_{0}$ is the vacuum wavelength of light.
For finding propagating modes we solve an 
eigenvalue problem for the propagation constant $k_{z}$ and propagation mode $\MyField{E}_{\mathrm{pm}}(x, y)$,
using the propagation mode solver included in our programme package {\it JCMsuite}~\cite{Burger2008ipnra}.
Figure~\ref{fig_wg_mode} shows a computed field distribution for the physical parameters as defined in 
Section~\ref{section_setup}.
The mode is concentrated at a very small cross-section~\cite{Oulton2008njp}.
The computed effective refractive index for the given parameters is $n_{\mathrm{eff}}\sim 1.2374+0.0118i$. 
We have checked that we can compute this number to a high numerical accuracy with relative errors of the 
real and imaginary parts of below $10^{-7}$. 
Figure~\ref{fig_wg_mode_conv} shows the corresponding convergence plot: The relative error of $n_{\mathrm{eff}}$ 
is computed for different levels of grid refinement.
The higher the level of grid refinement, the higher is the number of unknowns in the FEM problem. 
The relative error of a measured quantity $\alpha(N)$ (here real and imaginary parts of $n_{\mathrm{eff}}$) 
is defined as $\Delta \alpha(N) = |\alpha(N)-\alpha_\mathrm{qe}|/|\alpha_\mathrm{qe}|$, where 
$\alpha_\mathrm{qe}$ is the quantity computed from the quasi-exact solution, i.e., 
from a solution on a finer mesh than the meshes of the solutions corresponding to $\alpha(N)$. 
It would be desirable to compare $\alpha(N)$ to an analytical solution, but
for problems where an analytical solution is not available, the quasi-exact solution 
is used as a makeshift. 
In previous works we also have checked that our methods converge to analytical results~\cite{Hoffmann2009a} 
and -- within the various numerical uncertainties -- to 
the same numerical results as other software implementations of FEM and of other rigorous methods 
for solving Maxwell's equations~\cite{Hoffmann2009a,Lockau2009a}.
Here, we have used finite elements of second polynomial order; 
typical computation times on a standard PC are in the range of seconds to minutes for the plotted results.  

\begin{figure}[b]
\psfrag{x}{\sffamily $x$}
\psfrag{y}{\sffamily $y$}
\psfrag{z}{\sffamily $z$}
\psfrag{L}{\sffamily $L$}
\begin{center}
  \includegraphics[width=.98\textwidth]{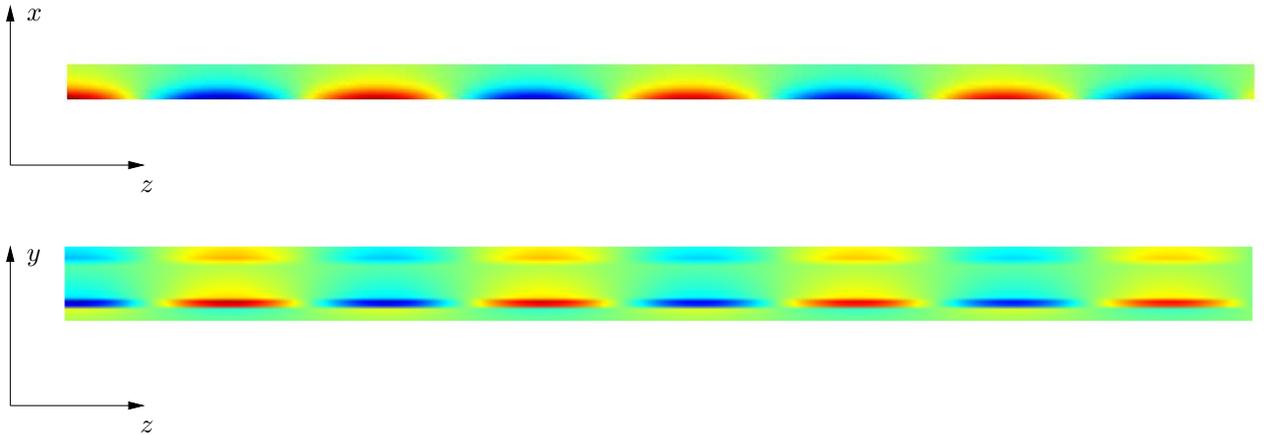}
  \caption{
Field distribution of a waveguide mode propagating through the 3D plasmonic waveguide. 
Top: $xz$-cross-section, $y$-component of the electric field in a pseudocolor visualization. 
Bottom: $yz$-cross-section, $y$-component of the electric field in a pseudocolor visualization. 
}
\label{fig_wg_3d_field}
\end{center}
\end{figure}
\subsection{3D simulation of light propagation over a finite distance of an infinite plasmonic waveguide}

In this Section results on 3D light scattering simulations of mode propagation through 
the investigated waveguide are presented.  
We performed simulations of 
light propagation over a certain distance $L$ of the infinite waveguide. 
The eigenvalue computed with the propagating mode solver, as described in 
Section~\ref{section_2d}, provides a very accurate value of expected transmission $T$
over a certain distance. 
The transmission is given by 
$T=\exp(-4\pi L\Im (n_{\mathrm{eff}})/\lambda_0) $.
We can use this number to check the accuracy of numerical results for specific numerical parameter 
settings for the 3D setup.

\begin{figure}[h]
\psfrag{Relative error (T)}{\sffamily \qquad \quad $\Delta T$}
\psfrag{N}{\sffamily $N$}
\psfrag{Convergence: Transmission / L = 1500nm}{\sffamily }
\begin{center}
  \includegraphics[width=.48\textwidth]{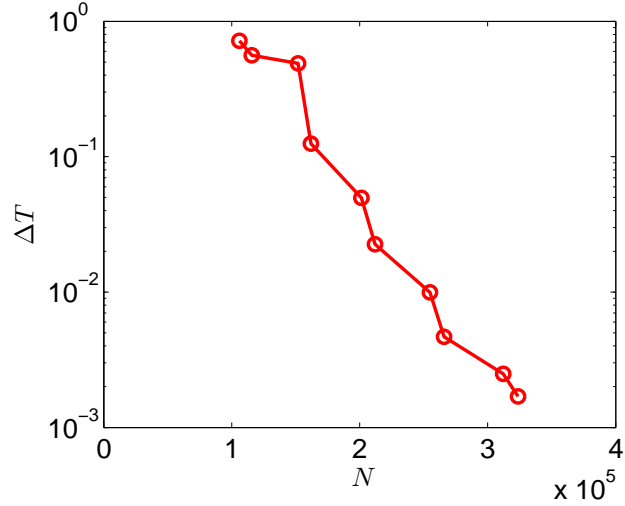}
  \caption{
Computation of light propagating through a 3D plasmonic waveguide. 
Convergence of the relative error of transmission, $\Delta T$, over a distance of 
$L=1.5\,\mu$m with number of unknowns of 
the FEM problem, $N$. 
}
\label{fig_wg_3d_conv}
\end{center}
\end{figure}
 
For simulating transmission through the waveguide at specific wavelength $\lambda_0$, 
we first compute the fundamental 
propagation mode of the waveguide at $\lambda_0$ 
as described in Section~\ref{section_2d}.
The obtained mode field is applied as input data to one of the boundaries of the 3D 
computational domain (front boundary in Fig.~\ref{fig_schema_pl}), such that the mode propagates 
in $+z$-direction.
We then compute the scattered light field in the setup corresponding to this excitation 
using higher-order finite-elements. 
A typical 3D computational domain is depicted in  Fig.~\ref{fig_schema_pl}. 
Transparent boundary conditions take into account the specific geometry 
of the problem where waveguides are modelled to  
extend to infinity in the exterior domain~\cite{Zschiedrich2006pml}. 
In post-processes we extract energy fluxes through interfaces and field distributions 
in several cross-sections from the 3D light field distribution.
The transmission is given by the quotient of transmitted and incoming energy flux, $T = E_\mathrm{t}/E_\mathrm{in}$.

Figure~\ref{fig_wg_3d_field} visualizes the computed electric field intensity in a cross-section 
in a $xz$-plane in the center of the MgF$_2$ layer (top) and in a cross-section 
in a $yz$-plane at $x=0$ (through the center of the nanowire, bottom). 
Please note the slight decay of the amplitude by a factor of about 0.8 over the propagation distance of 
$L=1.5\,\mu$m.
Figure~\ref{fig_wg_3d_conv} shows how the transmission through the waveguide computed with 3D 
light scattering simulations converges to the quasi-exact result obtained from the 
2D propagation mode computation. Accuracies in the range of 0.1\%
are easily reached. Computation times for the results shown in this plot 
range between one and few minutes on a standard workstation with RAM below 10GB. 

\begin{figure}[h]
\centering
\psfrag{Transmission}{\sffamily \qquad $T$}
\psfrag{Cavity length [nm]}{\sffamily \qquad \quad $L$ [nm]}
\psfrag{W = 100  nm}{\sffamily W\,=\,100\,nm}
\psfrag{W = 150  nm}{\sffamily W\,=\,150\,nm}
\psfrag{W = 200  nm}{\sffamily W\,=\,200\,nm}
\psfrag{L}{\sffamily $L$}
\psfrag{W}{\sffamily $W$}
\psfrag{d1}{\sffamily $h_1$}
\psfrag{d2}{\sffamily $D$}
\psfrag{x}{\sffamily $x$}
\psfrag{y}{\sffamily $y$}
\psfrag{z}{\sffamily $z$}
\includegraphics[width=0.5\textwidth]{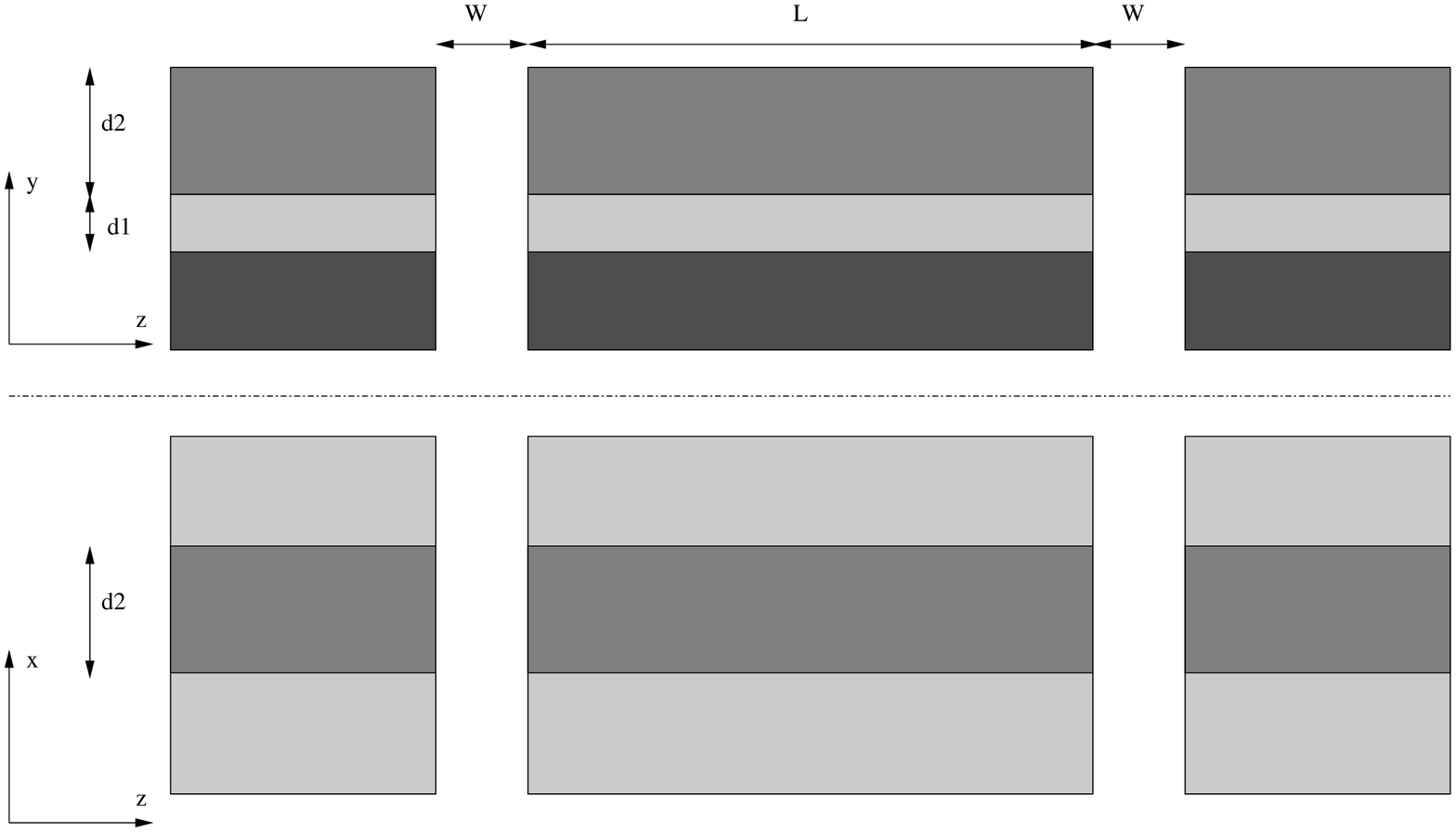}
\includegraphics[width=.43\textwidth]{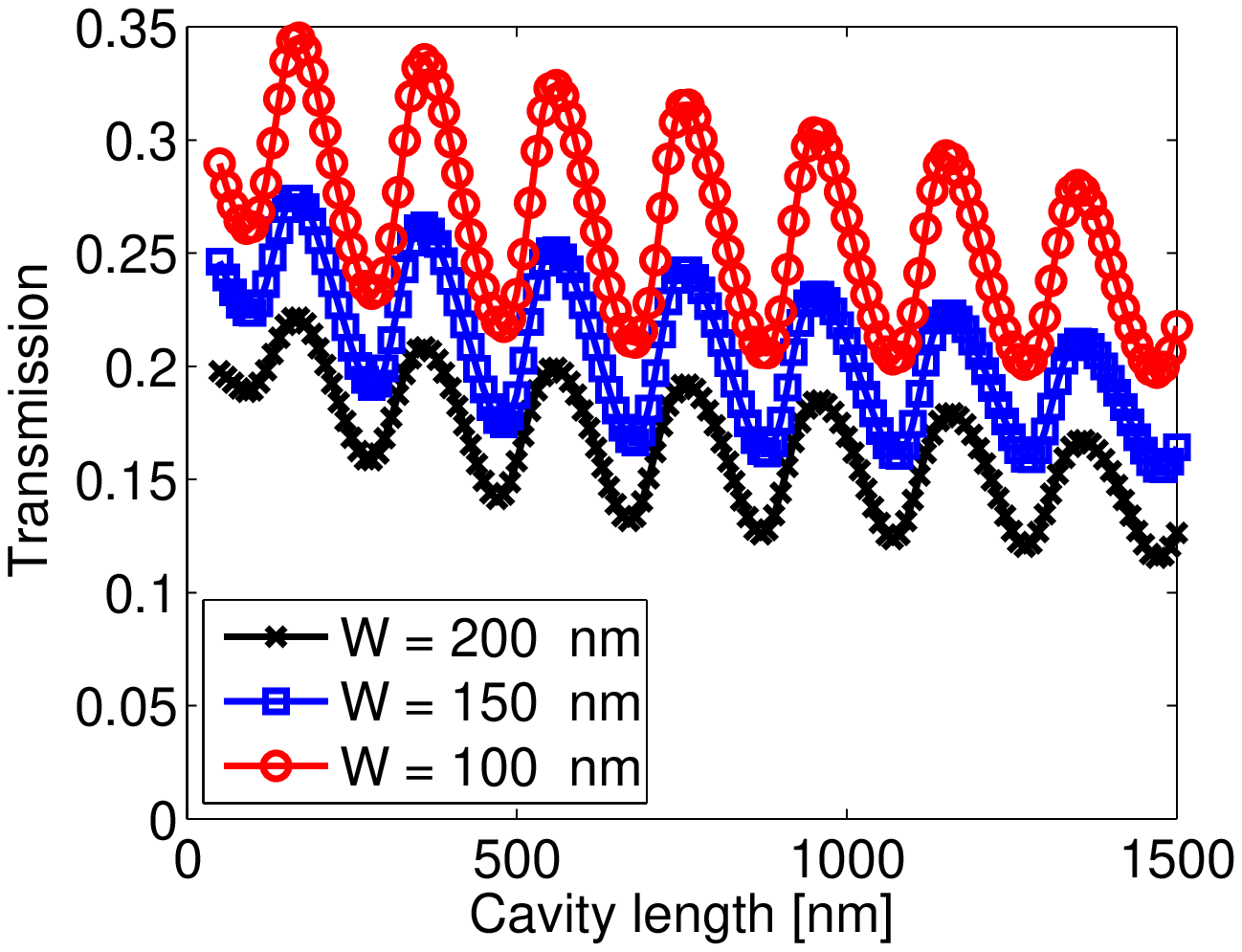}
\caption{Left: Schematic of the setup (top view and sideview): 
CdS nanowire of diameter $D$ and length $L$ placed on a thin layer of 
MgF$_2$ ($h_1=5\,$nm) on a silver layer (subspace). 
Coupling to the nanowire by identical, semi-infinite waveguides on either side of the 
device, separated by a distance $W$.
Right: Transmission $T$ through plasmonic waveguides of varied 
length $L$, separated from incoupler and outcoupler
waveguides by distances $W=100, 150, 200\,$nm. 
Vacuum wavelength of the incoming waveguide modes is 489.0\,nm.
}
\label{fig_finite_schema_graph}
\end{figure}

\subsection{3D simulation of transmission through a finite plasmonic waveguide}
\label{section_transmission_finite}
\begin{figure}[b]
\psfrag{x}{\sffamily $x$}
\psfrag{y}{\sffamily $y$}
\psfrag{z}{\sffamily $z$}
\psfrag{L}{\sffamily $L$}
\psfrag{N}{\sffamily $N$}
\begin{center}
\includegraphics[width=.98\textwidth]{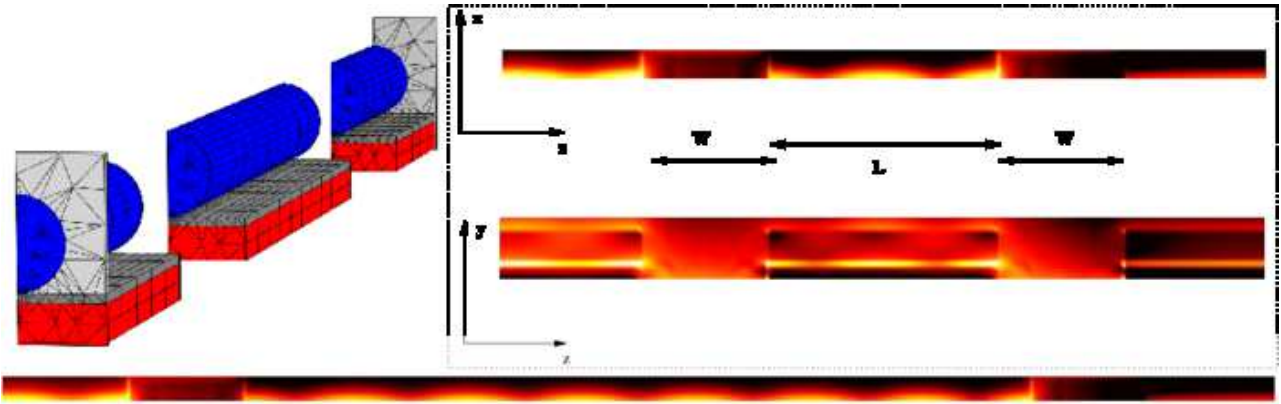}
  \caption{
Top left: Mesh of the 3D setup. 
Top right: Field distribution of a resonance mode for $L=370nm$ and $W=200\,$nm: 
$xz$-and $yz$-cross-sections, electric field intensity in a pseudocolor visualization (logarithmic scale). 
Bottom: $L=1360nm$, $W=200\,$nm, $xz$-cross-section, electric field intensity in a pseudocolor visualization (logarithmic scale). 
}
\label{fig_finite_mesh_field}
\end{center}
\end{figure}

For investigating resonances of a finite plasmonic waveguide we performed a numerical experiment 
where we simulate transmission through such a device. 
Figure~\ref{fig_finite_schema_graph} shows a schematic of the setup, 
a corresponding FEM mesh is shown in Figure~\ref{fig_finite_mesh_field} (left). 
Coupling to the finite waveguide and outcoupling from the waveguide is done by identical, 
semi-infinite plasmonic waveguides, which are separated from the finite waveguide by a 
distance $W$. The total length of the computational domain in these simulations is fixed 
to 2.2\,$\mu$m.
Figure~\ref{fig_finite_schema_graph} (right) shows transmission spectra through the plasmonic 
waveguide resonators.
In these spectra the resonator length $L$ has been varied at fixed 
vacuum wavelength of the incoming waveguide mode of 489.0\,nm.
We have recorded spectra for various widths $W$ of the gaps separating the resonator from the 
in- and outcoupling waveguides. 
Clearly, resonance peaks are observed in the transmission spectra. 
As expected, for increasing gap width $W$, the overall transmission decreases. 
Also resonance widths and positions slightly shift with the mirror width $W$. 
Figure~\ref{fig_finite_mesh_field}  shows field distributions in cross-sections through 
the 3D computational domain for $W=200\,$nm and $L=370\,$nm, and for 
$W=200\,$nm and $L=1360\,$nm, both at a transmission maximum.
High intensities at the end facets of the resonator can be observed ({\it cf.} a comparable 
experimental image by Oulton {\it et al.}~\cite{Oulton2009n}), corresponding to a standing-wave 
excitation of the resonator.

\begin{figure}[t]
\psfrag{Resonance wavelength [nm]}{\sffamily \qquad \quad $\lambda_0$ [nm]}
\psfrag{Cavity length [nm]}{\sffamily \qquad $L$ [nm]}
\psfrag{Q-factor}{\sffamily \quad $Q$}
\begin{center}
  \includegraphics[width=.48\textwidth]{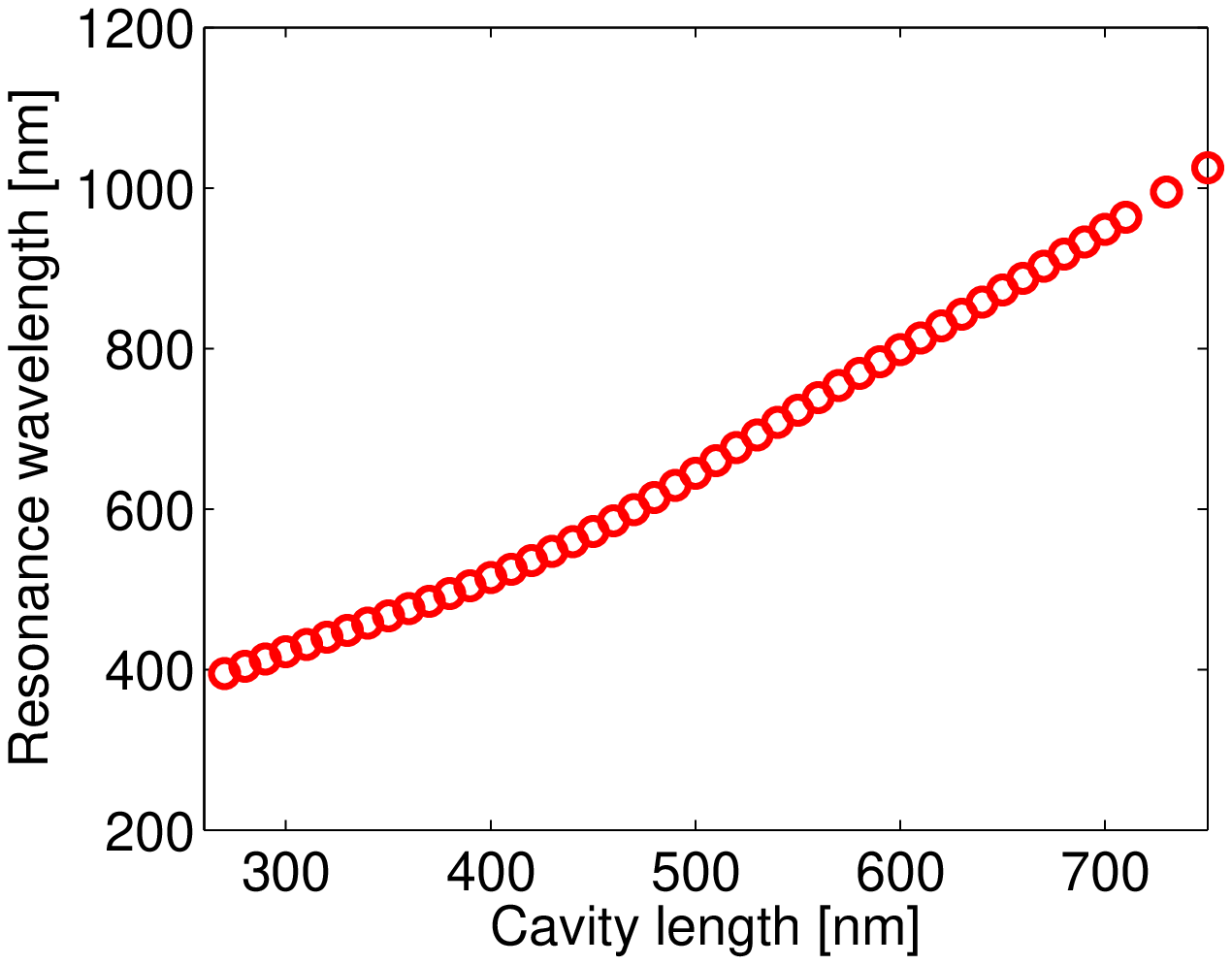}
  \includegraphics[width=.48\textwidth]{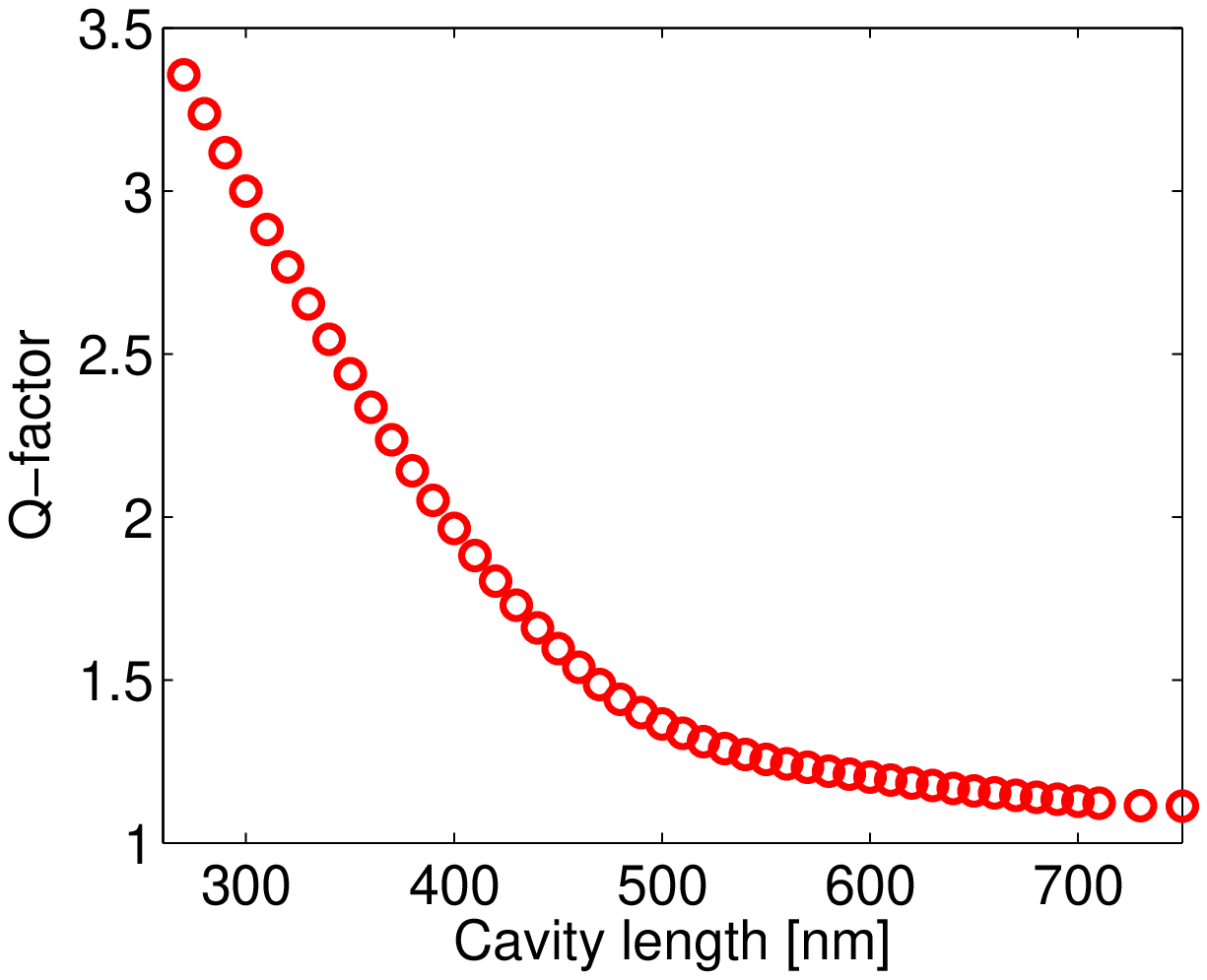}
  \caption{
Left: Dependence of resonance wavelength $\lambda_{0}$ on cavity length $L$.
Right: Dependence of cavity Q-factor on $L$.
}
\label{fig_res_wl_q}
\end{center}
\end{figure}

\begin{figure}[b]
\psfrag{x}{\sffamily $x$}
\psfrag{y}{\sffamily $y$}
\psfrag{z}{\sffamily $z$}
\psfrag{L}{\sffamily $L$}
\psfrag{N}{\sffamily $N$}
\psfrag{Real(omega)}{\sffamily $\Re (\omega)$}
\psfrag{Imag(omega)}{\sffamily $\Im (\omega)$}
\psfrag{Relative error}{\sffamily $\Delta\Re(\omega) / \Delta\Im(\omega)$}
\begin{center}
  \includegraphics[width=.56\textwidth]{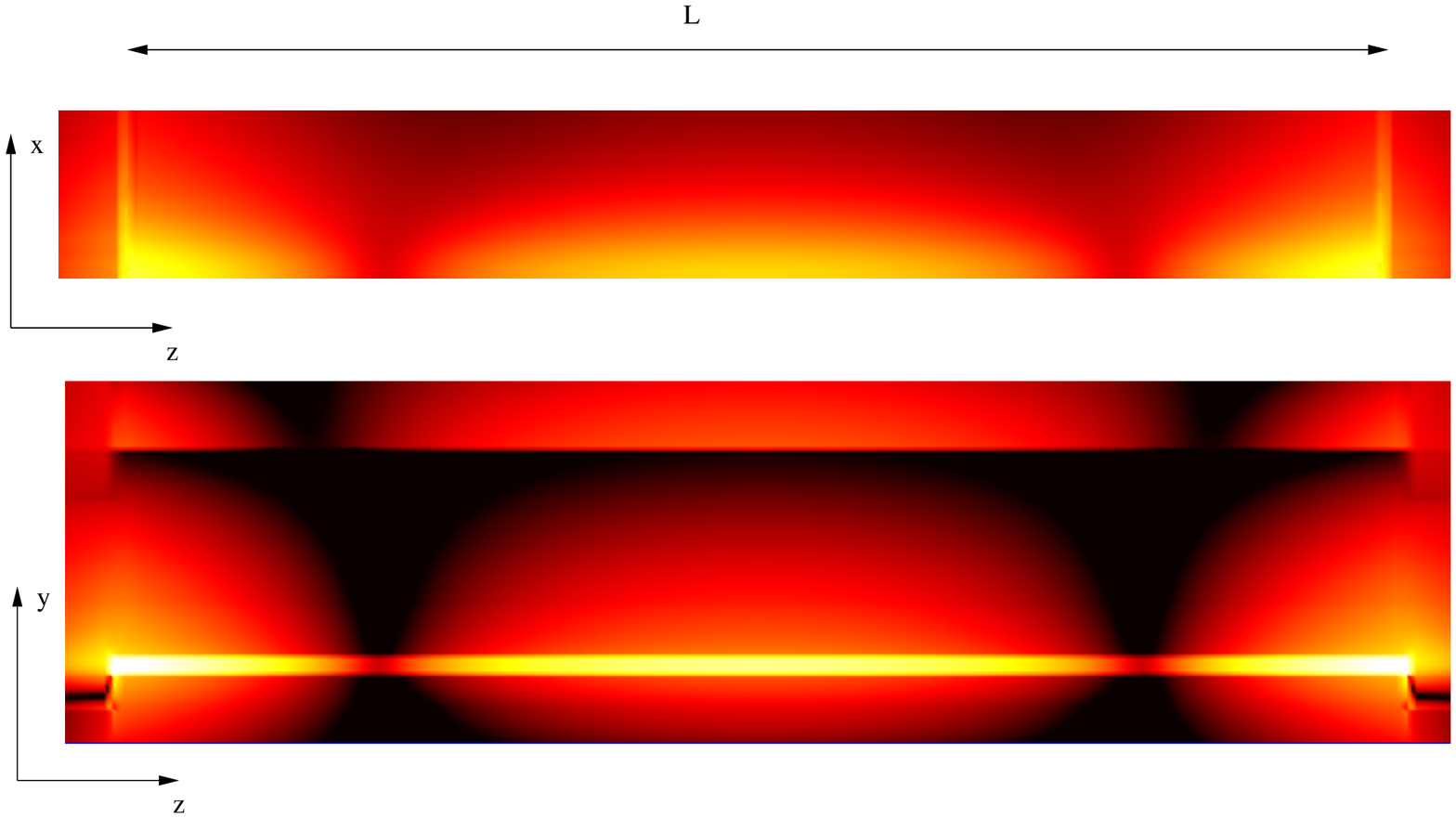}
  \includegraphics[width=.40\textwidth]{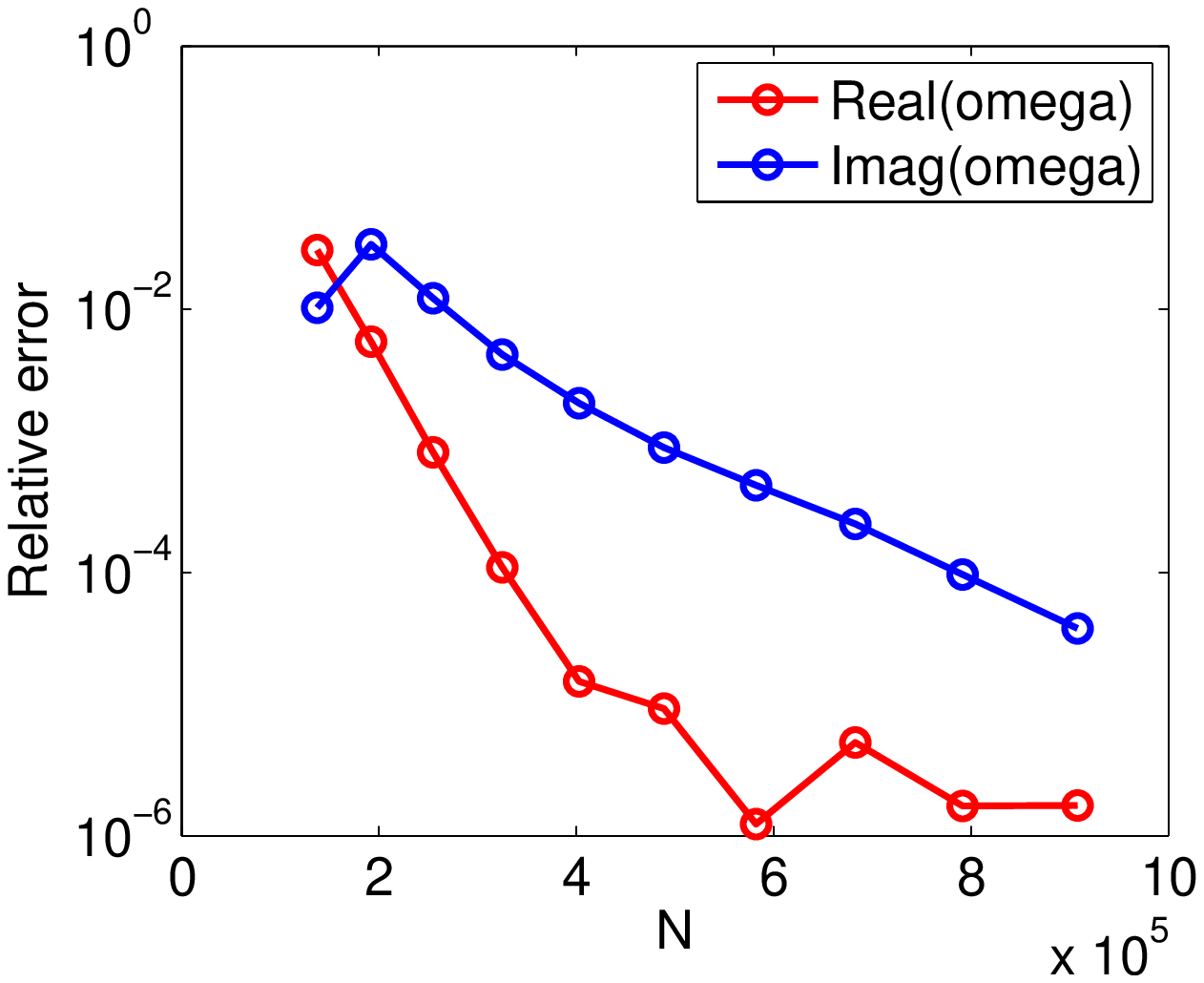}
  \caption{
Left: Field distribution of a resonance mode for $L=374nm$ and $\lambda_{0}\sim 489\,$nm. 
Top: $xz$-cross-section, $y$-component of the electric field in a pseudocolor visualization (logarithmic scale). 
Bottom: $yz$-cross-section, $y$-component of the electric field in a pseudocolor visualization (logarithmic scale). 
Right: Numerical convergence of the simulated eigenvalues: 
Dependence of the relative errors of the real and imaginary part of the complex eigenfrequency
on the number of unknowns $N$ of the FEM problem.  
}
\label{fig_res_conv}
\end{center}
\end{figure}
\subsection{Direct simulation of resonances of a finite plasmonic waveguide}
As has been experimentally demonstrated~\cite{Oulton2009n}, a plasmonic waveguide of finite 
length can be used as nanolaser cavity. 
Computing Q-factor and resonance wavelength of such a resonator from transmission spectra of a resonator, 
as obtained from light-scattering simulations (see Section~\ref{section_transmission_finite}) can be  
time-consuming, and -- more importantly -- the scattering response of a low-Q resonator makes 
quantitative investigations of the optical properties difficult due to the presence of a strong 
diffuse scattered field. 
For computing optical properties of such resonator we therefore use a FEM eigenmode solver included 
in {\it JCMsuite}. 
We have recently shown that this solver allows to compute resonance wavelength 
and Q-factor for high-Q photonic crystal microcavities at very high accuracy~\cite{Burger2009Tacona}. 
Given a geometrical setup with specific material parameters,
one computes an electric field distribution $E$
and a complex eigenfrequency $\omega$ which satisfy Maxwell's time-harmonic wave equation 
$$\nabla \times \mu^{-1} \nabla \times E = \omega^2 \varepsilon E$$
without sources; 
electric permittivity and magnetic permeability are denoted by $\varepsilon$ and $\mu$, respectively.
%Transparent boundary conditions take into account the specific geometry 
%of the problem where waveguides are modelled to  
%extend to infinity in the exterior domain~\cite{Zschiedrich2006pml}. 
When the eigenmode ($E, \omega$) is computed, the respective $Q$-factor is deduced from the real and imaginary parts 
of the complex eigenfrequency,
$Q=\Re(\omega)/(-2\Im(\omega)),$
the resonance wavelength $\lambda_{0}$ is given by $\lambda_{0}=2\pi c_0/\Re(\omega)$, with the speed of light $c_0$.

In order to find the cavity length $L$ for which the cavity is resonant at the target wavelength
$\lambda_{0}=489\,$nm, we have performed simulations with varied $L$ and otherwise fixed parameters. 
The resulting plots on the behavior of
resonance wavelength and Q-factor change with cavity length are shown in Figure~\ref{fig_res_wl_q}. 
For these results, we have again chosen physical parameters as defined in Section~\ref{section_setup}.
As can be seen from the Figure, a resonance wavelength of $\lambda_{0}=489\,$nm is obtained for a 
nanowire length of $L=374\,$nm 
with a corresponding Q-factor of $Q\sim 2.2$. 
This is in very good agreement with the results of Section~\ref{section_transmission_finite}.
Cross-sections through the 3D field distribution visualizing the resonant electric field distribution 
at $L=374\,$nm are shown in Figure~\ref{fig_res_conv} (left).
We have performed a convergence study for this specific cavity length. Here we investigate the convergence of 
the real and of the imaginary part of the complex eigenfrequency. For fixed geometrical parameters and for 
a finite element polynomial degree of $p=2$ we increase the spatial resolution of the FEM mesh, leading to 
finer and finer meshes and FEM problems with increasing number of unknowns of the algebraic problem, $N$. 
Figure~\ref{fig_res_conv} shows that for this specific setting, both, $\lambda_{0}$ and $Q$ 
can be computed to a high numerical 
accuracy with a relative error below $10^{-4}$.

\section{Conclusion}
We have presented a numerical method for investigating nanoplasmonic devices like waveguides and 
cavities. 
Waveguide mode computations, 3D light scattering computations and 3D resonance mode computations have 
been performed. 
Accurate results have been obtained using higher-order finite elements and adaptive grid refinement. 

\bibliography{/home/numerik/bzfburge/texte/biblios/phcbibli,/home/numerik/bzfburge/texte/biblios/group,/home/numerik/bzfburge/texte/biblios/lithography}

\begin{thebibliography}{10}

\bibitem{Bergman2003prl}
Bergman, D.~J. and Stockman, M.~I., ``Surface plasmon amplification by
  stimulated emission of radiation: Quantum generation of coherent surface
  plasmons in nanosystems,'' {\em Phys. Rev. Lett.}~{\bf 90}(2),  027402
  (2003).

\bibitem{Anker2008nm}
Anker, J.~N., Hall, W.~P., Lyandres, O., Shah, N.~C., Zhao, J., and Duyne, R.
  P.~V., ``Biosensing with plasmonic nanosensors,'' {\em Nature Materials}~{\bf
  7},  442 -- 453 (2008).

\bibitem{Noginov2009nature}
Noginov, M.~A., Zhu, G., Belgrave, A.~M., Bakker, R., Shalaev, V.~M.,
  Narimanov, E.~E., Stout, S., Herz, E., Suteewong, T., and Wiesner, U.,
  ``Demonstration of a spaser-based nanolaser,'' {\em Nature}~{\bf 460},  1110
  (2009).

\bibitem{Hill2009oe}
Hill, M.~T., Marell, M., Leong, E. S.~P., Smalbrugge, B., Zhu, Y., Sun, M., van
  Veldhoven, P.~J., Geluk, E.~J., Karouta, F., Oei, Y.-S., N\"{o}tzel, R.,
  Ning, C.-Z., and Smit, M.~K., ``Lasing in metal-insulator-metal
  sub-wavelength plasmonic waveguides,'' {\em Opt. Express}~{\bf 17}(13),
  11107--11112 (2009).

\bibitem{Oulton2009n}
Oulton, R.~F., Sorger, V.~J., Zentgraf, T., Ma, R.-M., Gladden, C., Dai, L.,
  Bartal, G., and Zhang, X., ``Plasmon lasers at deep subwavelength scale,''
  {\em Nature}~{\bf 461},  629 (2009).

\bibitem{Chang2008oe}
Chang, S.-W., Ni, C.-Y.~A., and Chuang, S.-L., ``Theory for bowtie plasmonic
  nanolasers,'' {\em Opt. Express}~{\bf 16}(14),  10580--10595 (2008).

\bibitem{Oulton2008njp}
Oulton, R.~F., Bartal, G., Pile, D. F.~P., and Zhang, X., ``Confinement and
  propagation characteristics of subwavelength plasmonic modes,'' {\em New
  Journal of Physics}~{\bf 10}(10),  105018 (14pp) (2008).

\bibitem{Oulton2008np}
Oulton, R.~F., Sorger, V.~J., Genov, D.~A., Pile, D. F.~P., and Zhang, X., ``A
  hybrid plasmonic waveguide for subwavelength confinement and long-range
  propagation,'' {\em Nature Photonics}~{\bf 2},  496 (2008).

\bibitem{Hoffmann2009a}
Hoffmann, J., Hafner, C., Leidenberger, P., Hesselbarth, J., and Burger, S.,
  ``Comparison of electromagnetic field solvers for the 3d analysis of
  plasmonic nano antennas,'' in [{\em Modeling Aspects in Optical
  Metrology}{\nolinebreak\hspace{0.1em}]},  Bosse, H. and Bodermann, B., eds.,
  {\bf 7390},  73900J, Proc. SPIE (2009).

\bibitem{Lockau2009a}
Lockau, D., Zschiedrich, L., and Burger, S., ``Accurate simulation of light
  transmission through subwavelength apertures in metal films,'' {\em J. Opt.
  A: Pure Appl. Opt.}~{\bf 11},  114013 (2009).

\bibitem{Tyagi2008oe}
Tyagi, H.~K., Schmidt, M.~A., Sempere, L.~P., and {Russell, P.\,St.\,J.},
  ``Optical properties of photonic crystal fiber with integral micron-sized
  {Ge} wire,'' {\em Opt. Express}~{\bf 16}(22),  17227--17236 (2008).

\bibitem{Zschiedrich2006pml}
Zschiedrich, L., Klose, R., Sch\"adle, A., and Schmidt, F., ``A new finite
  element realization of the {P}erfectly {M}atched {L}ayer {M}ethod for
  {H}elmholtz scattering problems on polygonal domains in 2{D},'' {\em J.
  Comput. Appl. Math.}~{\bf 188},  12--32 (2006).

\bibitem{Burger2008ipnra}
Burger, S., Zschiedrich, L., Pomplun, J., and Schmidt, F., ``{JCMsuite}: {A}n
  adaptive {FEM} solver for precise simulations in nano-optics,'' in [{\em
  Integrated Photonics and Nanophotonics Research and
  Applications}{\nolinebreak\hspace{0.1em}]},   ITuE4, Optical Society of
  America (2008).

\bibitem{Burger2009Tacona}
Burger, S. and Zschiedrich, L., ``Numerical investigation of
  silicon-on-insulator {1D} photonic crystal microcavities,'' in [{\em
  Theoretical and computational nanooptics: Proceedings of the 2nd
  International Workshop}{\nolinebreak\hspace{0.1em}]},  Chigrin, D.~N., ed.,
  {\bf 1176},  43--45, AIP (2009).

\end{thebibliography}
\bibliographystyle{spiebib}  

\end{document}